\documentclass[12pt]{iopart}
\usepackage{epsfig}
\usepackage{amssymb}
\usepackage{graphicx}
\usepackage{amssymb}
\usepackage{color}

\begin{document}

\title{Electronic Structure of an Iron-Porphyrin Derivative on Au(111)}

\author{Arnulf Stein$^1$, Daniela Rolf$^2$, Christian Lotze$^2$, Constantin Czekelius$^3$, Katharina J. Franke$^2$, and Petra Tegeder$^1$}

\address{$^1$Ruprecht-Karls-Universit\"{a}t Heidelberg, Physikalisch-Chemisches Institut, Im Neuenheimer Feld 253, 69120 Heidelberg, Germany}
\address{$^2$Freie Universit\"{a}t Berlin, Fachbereich Physik, Arnimallee 14, 14195 Berlin, Germany}
\address{$^3$Heinrich-Heine-Universit\"at D\"usseldorf, Institut f\"ur Organische Chemie und Makromolekulare Chemie, Universit\"atsstra\ss e 1, 40225 D\"usseldorf, Germany}
\ead{tegeder@uni-heidelberg.de}
\vspace{10pt}
%\begin{indented}
%\item[]November 19, 2018
%\end{indented}

\begin{abstract}
Surface-bound porphyrins are promising candidates for molecular switches, electronics and spintronics. Here, we studied the structural and the electronic properties of
Fe-tetra-pyridil-porphyrin adsorbed on Au(111) in the monolayer regime. We combined scanning tunneling
microscopy/spectroscopy, ultraviolet photoemission, and two-photon photoemission to determine the energy levels of the frontier molecular orbitals. We also resolved an excitonic state with a binding energy of 420 meV, which allowed us to compare the electronic transport gap with the optical gap.

\end{abstract}

%\maketitle

\section{Introduction}
Porphyrin molecules play an important role in many
chemical and biological processes. For instance, they are the essential part in cytochrome C, where an incident photon excites the porphyrin, which in turn triggers an electron transfer process along a complex molecular pathway. The flexibility of porphyrins and their electronic properties have also motivated their investigation as building blocks in organic electronics. In potential devices, such as organic thin film transistors, light emitting diodes or photovoltaics, the electronic properties and their performance are strongly influenced by the interface to inorganic (metal) electrodes \cite{Otsuki2010, Jurow2010, Gottfried2015, Auwarter2015, Kesters2015, Marbach2015,Heinrich2015, Bockmann2016, carmen_rubio_verdu_orbital_selective_2017, Karan2018}.
The interfacial electronic structure is controlled by the adsorption geometry of the porphyrin molecules, which in turn is governed by the strength of lateral and vertical interactions. Due to the flexibility of the porphyrins many different molecular arrangements can be found on surfaces, where the molecules undergo intramolecular distortions in response to the external forces driving the self-assembly. These
properties of various porphyrins have been studied in detail using scanning tunneling microscopy (STM) \cite{Otsuki2010, Gottfried2015, Auwarter2015, Marbach2015}. However, information about excitonic states are rather limited. By means of ultraviolet photoemission (UPS) and inverse photoemission spectroscopy (IPS) the energy level alignment of a few porphyrin/substrate systems have been determined \cite{Giovanellia2017, Ruggieri2016, Rangan2016, Bussetti2016, Rangan2012, Cook2011, Rangan2010, Rojas2010}. However,  IPS is limited with respect to the energy resolution and the combination of UPS/IPS does not give insights into optical gaps and accordingly exciton binding energies. Two-photon photoemission (2PPE) spectroscopy has been proven to be a powerful tool, which provides the energetic position of occupied and unoccupied molecular electronic states (transport levels) but also excitonic states (optical gaps) and therefore exciton binding energies \cite{Zhu2004, Lindstrom2006, Yang2008, Zhu2014, Varene2011, Varene2012, Marks2013, Galbraith2014, Ilyas2014, Bogner2016, Bogner2015, Tegeder2017}. So far, only two 2PPE studies concerning porphyrin/substrate are available in literature \cite{Tognolini2015, Classen2017}. In the first case, the electronic structure of Mg tetraphenylporphyrin  films of 5 and 19 monolayer (ML) thickness on a Ag(100) substrate has been studied \cite{Classen2017} and the energetic positions of the lowest unoccupied molecular orbital (LUMO) and the highest occupied molecular orbital (HOMO) have been determined. In the second case, time-resolved 2PPE has been utilized to investigate the electron dynamics of unoccupied molecular states, which depends strongly on the substrate orientation \cite{Tognolini2015}.

\begin{figure}[hbt]
\centering \resizebox{0.3\hsize}{!}{\includegraphics*{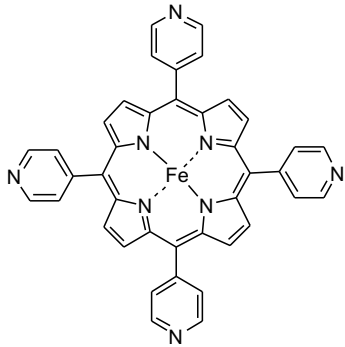}}
\caption{The Fe-5,10,15,20-tetra-pyridil-porphyrin (FeTPyP) molecule.}
\label{figure1}
\end{figure}

In this study, we investigate the adsorption and the electronic structure of Fe-5,10,15,20-tetra-pyridil-porphyrin (FeTPyP, see Fig. \ref{figure1}) adsorbed on Au(111) using STM, scanning tunneling spectroscopy (STS), UPS and 2PPE. We obtained a detailed picture on the energetic
positions of several affinity levels and ionization potentials originating
from the LUMOs and HOMOs in FeTPyP/Au(111). We identified the transport gap to be
2.2 eV as well as an exciton binding energy of 0.42 eV. Thus, the optical gap in FeTPyP adsorbed on Au(111) is 1.78 eV.

\section{Experimental Methods}
STM/STS and UPS/2PPE measurements were conducted in two
different ultrahigh vacuum chambers.
In both cases, a clean Au(111) substrate was prepared by a standard procedures of sputtering-annealing cycles under ultra-high vacuum conditions. FeTPyP-Cl
molecules were deposited from an effusion cell held at a temperature of 680 K while the surface was kept at 300 K. Upon adsorption, the molecules loose their Cl ligand (see below).

\textbf{STM/STS Experiments:} The as-prepared samples were pre-cooled and transferred into the STM, which is operated a 4.5 K. All images were taken in constant-current mode, whereas the dI/dV spectra were recorded at constant height with the feedback opened at the conductance set points as specified in the figure caption. A lock-in amplifier was used with a modulation amplitude of 10 mV.

\textbf{2PPE Experiments:} In 2PPE, a femtosecond pump laser pulse with a photon energy $h\nu_1$ induces electronic transitions from occupied electronic states of the substrate or the adsorbate into unoccupied electronic or virtual states. These excited states are then probed with a second laser pulse with photon energy $h\nu_2$ by lifting the excited electron above the vacuum level (E$_{vac}$). The kinetic energy of the emitted electrons is measured with a time-of-flight spectrometer. Photoelectron signals  can arise from occupied initial, unoccupied intermediate or final electronic states. Therefore photon energy dependent measurements are needed for the assignment (for details see Refs. \cite{bronner12, bronner11}). In a so-called one-color photoemission experiment $h\nu_1$ and $h\nu_2$ have the same energy, in a two-color experiment the energy of $h\nu_2$  is twice the energy of $h\nu_1$  (frequency-doubled). The UPS and 2PPE measurements have been performed at a substrate temperature of 90 K.
The FeTPyP coverage has been determined \emph{via} temperature-programmed desorption and work function measurements.

\section{Results and Discussion}
The electronic
structure of organic molecules adsorbed on metal surfaces essentially
depends on the electronic coupling between molecular  and
metal states and is a result of the adsorption geometry
of the adsorbate. Therefore, we first analyzed the adsorption structure of
 FeTPyP on Au(111) using STM.

\begin{figure}[hbt]
\centering \resizebox{0.7\hsize}{!}{\includegraphics*{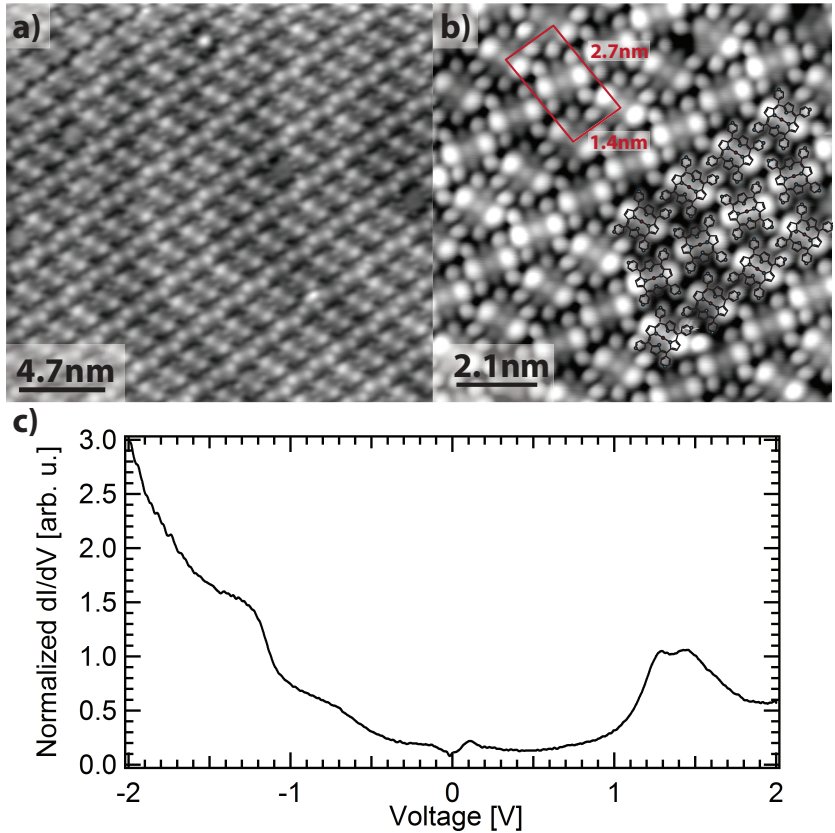}}
\caption{a) STM topography showing the formation of large, ordered islands of the FeTPyP molecules on the Au(111) surface. b) Close-up view of the arrangement of the molecules in alternating rows. The red rectangle depicts the unit cell. c) dI/dV spectrum recorded in the center of an FeTPyP molecule. The double resonance at 1.3 V and 1.45 V can be assigned to the LUMO and the LUMO+1 of the molecule, respectively. The resonance at -1.3 V is attributed to a lower lying HOMO-n (see text). Feedback opened at 2 V, 1 nA with $V_{mod}$ = 10mV.}
\label{figure2}
\end{figure}
An STM topography of the FeTPyP molecules on the Au(111) surface is shown in Fig. \ref{figure2} a). The molecules form large self-assembled islands that extend over several tens of nanometers. Underneath the molecule, the herringbone reconstruction of the Au surface is maintained, which indicates a rather weak interaction of the molecules with the substrate. In Fig. \ref{figure2} b), a close-up view of the molecular arrangement is depicted. The molecules arrange in two alternating rows, in which the molecules are aligned in parallel. Due to the saddle-shape distortion of the FeTPyP molecules upon adsorption \cite{auwarter_controlled_2007,auwarter_conformational_2007,zotti_ab_initio_2007, albrecht_direct_2016,wang_intramolecularly_2015, xianwen_chen_conformational_2017, liu_large_2017,carmen_rubio_verdu_orbital_selective_2017}, two of the pyrrole groups appear higher in the STM topography than the other two. We do not find any indication of a central ligand in the STM images, indicating the desorption of the Cl ligand upon adsorption on the metal surface. The desorption goes along with a reduction of the oxidation state of Fe from $III+$ to $II+$. This assignment of the oxidation state of Fe and the loss of the Cl ligand is corroborated by measurements of inelastic spin excitations, which revealed an integer spin state of S=1 or S=2 depending on the molecular assembly \cite{Rolf2018}.
Fig. \ref{figure2} c) shows a dI/dV spectrum recorded in the center of an FeTPyP molecule. At around 1.3 V and 1.45 V, a double-peak structure can be observed, which we assign to the LUMO and LUMO+1 of the molecule, respectively. At negative bias voltage the conductance increases around -0.6 V  and around -1.3 V a broad resonance is observed. At first glance one tends to assign this resonance to the HOMO. However, the overall observed density of states is very broad and it is difficult to discern individual resonances. STS is inherently less sensitive to occupied states since the tunneling barrier is increasing with increasing negative bias voltage. As we will demonstrate by means of UPS several occupied molecular electronic states are located in the energy region between -1.5 and -0.9 eV.

\begin{figure}[hbt]
\centering \resizebox{0.5\hsize}{!}{\includegraphics*{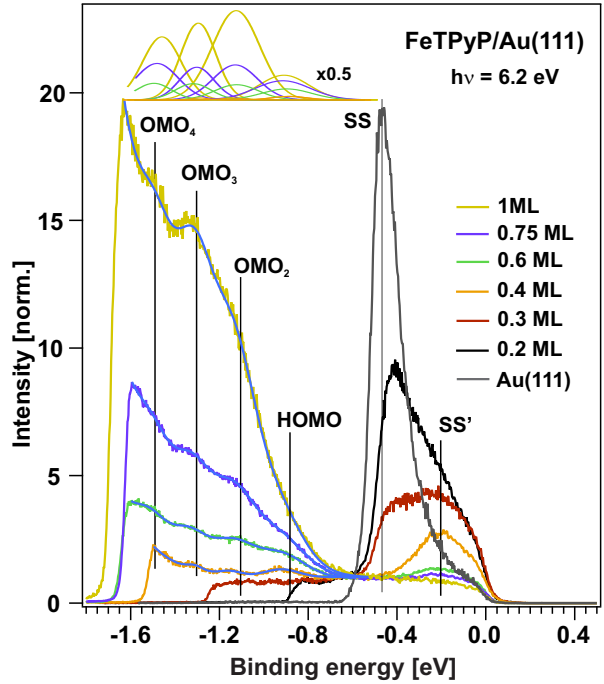}}
\caption{UPS data recorded at photon energy 6.2 eV as a function of FeTPyP coverage. OMO denotes an occupied molecular orbital, SS the Shockley surface state, and SS$^{'}$  the shifted surface state due to the adsorption of FeTPyP. The data have been fitted with an exponential background and Gaussian-shaped peaks.}
\label{figure3}
\end{figure}
For gaining complementary insight into the electronic structure, we utilized UPS and 2PPE spectroscopy. Figure \ref{figure3} shows UPS data on preparations with different sub-monolayer to monolayer coverages recorded with a photon energy of 6.2 eV. The spectrum of the bare Au(111) surface is dominated by the Shockley surfaces state (SS) located at -0.48 eV. Deposition of FeTPyP leads to a decrease of the work function and the SS loses intensity. In addition, around -0.2 eV the photoemission intensity increases, which is due to the adsorbate-induced shifted surface state (SS$^{'}$). Note that the work function decreases by 1 eV  from 5.5 eV for the bare Au(111) surface to 4.5 eV at a FeTPyP coverage of 1 monolayer (ML) as can be seen from the shift of the cut-off energy in the spectra.  Increasing the coverage above approximately 0.4 ML, FeTPyP-derived features start to dominate the spectrum (the photoemission intensity strongly rises with increasing coverage), the SS possesses nearly no intensity and at even higher coverages also the SS$^{'}$ intensity drops. The data have been fitted with an exponential background and Gaussian-shaped peaks to determine the energetic positions of the occupied molecular states. We assign the feature at -0.9 eV to photoemission from the HOMO. The three peaks at higher binding energies are attributed to lower lying occupied molecular orbitals (OMO) labeled as OMO$_{2}$, OMO$_{3}$, and OMO$_{4}$. They are located at -1.1, -1.3, and -1.5 eV. Thus, in contrast to the STS data, which exhibits a limited sensitivity to occupied states due to the increase in tunneling barrier height with increasing negative bias voltage, UPS enabled us to energetically resolve the occupied electronic structure of FeTPyP/Au(111) in more detail.

\begin{figure}[hbt]
\centering \resizebox{0.5\hsize}{!}{\includegraphics*{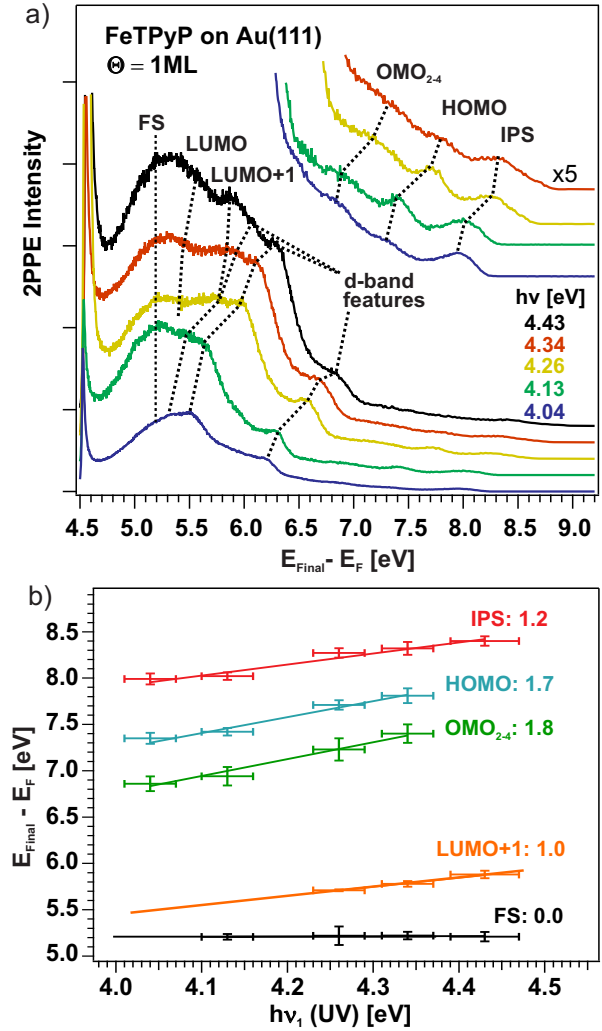}}
\caption{a) One-color 2PPE spectra of 1ML FeTPyP adsorbed on Au(111) recorded with different photon energies. The energy axis reveals the final state ($E_{Final}$) of the
photoemitted electrons with respect to $E_{F}$ ($E_{Final} - E_{F} = E_{kin} + \Phi$);
thus, the low-energy cutoff corresponds to the work function ($\Phi$) of the adsorbate-substrate system. The inset shows the high energy region, in which the intensity is enlarged by a factor of five. b) Photon energy-dependent peak
position to assign peaks observed in the 2PPE spectrum to occupied,
unoccupied intermediate or final electronic states. A slope of 1 suggests that a peak
originates from an unoccupied intermediate state, a slope of zero from an unoccupied final state (located above the vacuum level), while a slope of 2 is related to
peaks originating from occupied states. The energetic position of the final state is independent of the used photon energy, i.e., the kinetic energy is constant. This can be viewed as a resonant scattering event in which the photoexcited electron resides
transiently in the molecular resonance, followed by detachment and detection.}
\label{figure4}
\end{figure}
To elucidate also the energetic positions of unoccupied molecular and excitonic states of FeTPyP/Au(111) 2PPE has been utilized.  Figure \ref{figure4}a) shows one-color ($h\nu_1$ = $h\nu_2$) 2PPE spectra of 1 ML FeTPyP adsorbed on Au(111) recorded with different photon energies. The spectra show several peaks related to either occupied or unoccupied electronic states of the FeTPyP/Au(111)
systems.
On the basis of a detailed analysis of the photon energy-dependent measurement (see Fig. \ref{figure4}a)), we can identify whether the photoemission peaks originate
from occupied initial, unoccupied intermediate or final states in
the 2PPE process. A scaling of the final-state energy with the incident photon energy with a slope of zero indicates the presence of an unoccupied resonance above the vacuum level. In contrast, a slope of one indicates photoemission from an unoccupied intermediate state, while a slope of two signifies photoemission from an occupied state.

As expected, we observe the gold \emph{d}-bands according to $E_{Final}-E_{F}-2h\nu_1$ at -2.0, -2.6, and -2.9 eV  \cite{Eckardt1984, Courths1986, Hagen2008}.
In agreement with the UPS data, we detect photoemission peaks with a slope of about two in the photon energy-dependent spectra originating from the HOMO and other OMOs. The latter appear as a broad feature, not resolving  the contributions of the individual occupied states (OMO$_{2}$ - OMO$_{4}$). In addition,  we detect a peak without any dependence on the incident photon energy. This peak arises from photoemission from an unoccupied molecular final state (FS) at $E_{Final} - E_{F}$ = 5.18$\pm$0.08 eV above the Fermi level, and accordingly 0.68$\pm$0.07 eV above the vacuum level.
Moreover, we find a peak which scales linearly with incident photon energy. It stems from an intermediate state at 1.44$\pm$0.09 eV above $E_{F}$. We attribute this peak to photoemission from the  LUMO+1. This is in agreement with the STS measurement shown above. The LUMO, which has been found at 1.3 eV in the tunneling spectra is hardly visible in the spectra recorded with low photon energies. However, we will show that this peak can be better resolved by two-color (2$h\nu_1$ = $h\nu_2$) 2PPE data.
The photoemission peak close to the high-energy cut-off can be assigned to the first (\emph{n} = 1) image potential state (IPS) located 3.87$\pm$0.07 eV  above $E_{F}$. This corresponds to an energy of 0.57 eV below the vacuum level, and has been found for other adsorbate-covered Au(111) surfaces \cite{bronner11, Varene2011, Bronner2012, Gerbert2017, Gerbert2017a, Stein2017}.

 \begin{figure}[hbt]
\centering \resizebox{0.45\hsize}{!}{\includegraphics*{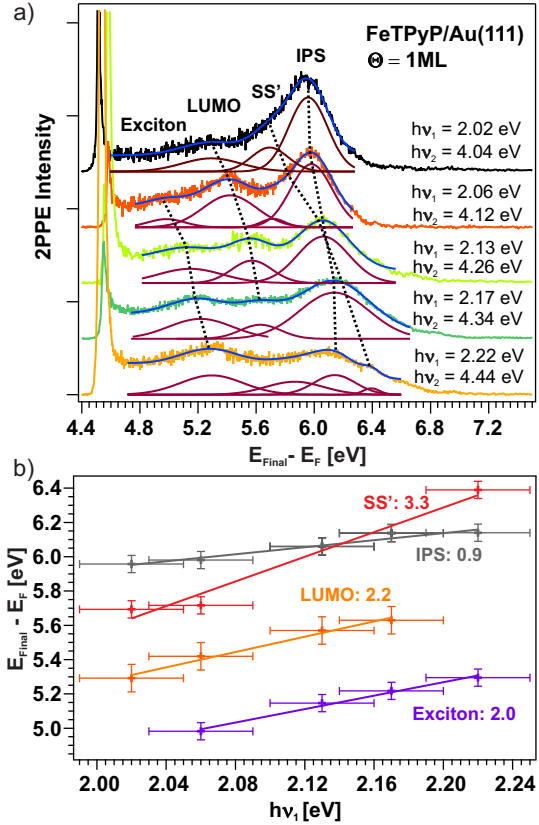}}
\caption{(a) Two-color (2$h\nu_1$ = $h\nu_2$) 2PPE
spectra of 1ML FeTPyP on Au(111) recorded with different photon energies. The data are fitted with an
exponential background and Gaussian-shaped peaks.
(b) Photon energy-dependent peak positions for the assignments of
the peaks observed in the 2PPE spectra to occupied or unoccupied
electronic states (slope = 1 and 2 for unoccupied states; slope = 3 for occupied
states).}
\label{figure5}
\end{figure}
Figure \ref{figure5} shows the two-color 2PPE spectra of 1 ML FeTPyP/Au(111) recorded with different photon energies and the corresponding analysis of the photon-energy dependent energetic position of the observed photoemission peaks. Apart from the SS$^{'}$, which has been observed in the UPS measurements (see Fig. \ref{figure3}) and the IPS also seen in the one-color 2PPE data (see Fig.\ref{figure4}), we found two additional features which arise from unoccupied molecular states. They are located at 1.29$\pm$0.07 eV and 0.87$\pm$0.1 eV above $E_{F}$, respectively. We associate the peak at 1.29$\pm$0.07 eV with the LUMO in accordance with the STS results (see Fig. \ref{figure2} c)) and the 2PPE feature at 0.87$\pm$0.1 eV (not seen in STS) to an excitonic state.

\begin{figure}[hbt]
\centering \resizebox{0.45\hsize}{!}{\includegraphics*{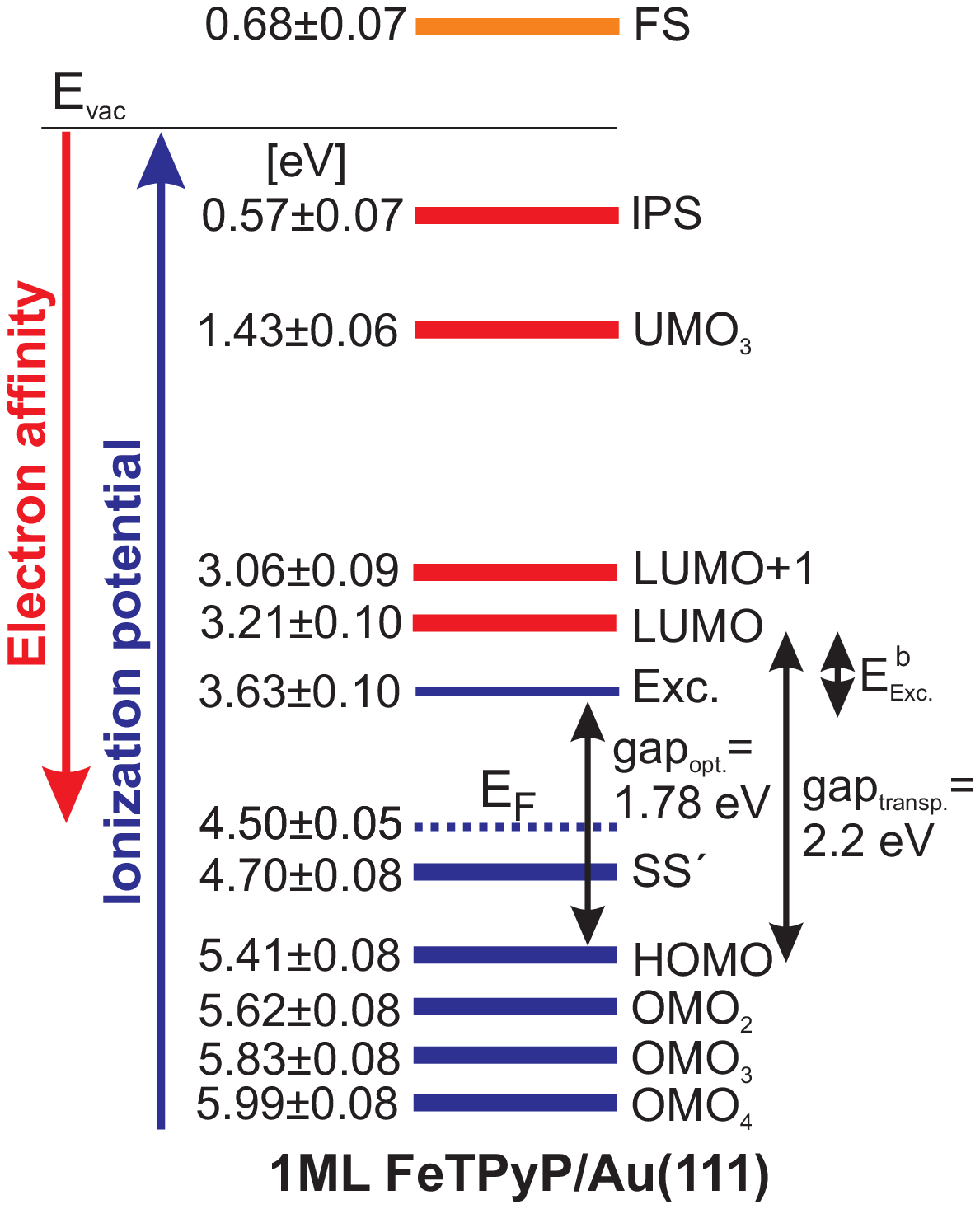}}
\caption{Energy level diagram of 1 ML FeTPyP/Au(111) with respect to the vacuum level ($E_{vac}$). The blue levels
are the ionization potentials and the red ones are the
electron affinities; $E_{F}$ is the Au(111) Fermi level; OMO is an occupied molecular orbital and UMO an unoccupied molecular orbital.}
\label{figure6}
\end{figure}

Combining the results from STS, UPS and 2PPE measurements, we arrive at a complete diagram of unoccupied as well as occupied molecular electronic states. In Fig. \ref{figure6}, we summarize the level alignments with respect to the vacuum
level ($E_{vac}$).
As discussed previously \cite{Bogner2016, Bogner2015, Bronner2016}, one has to consider that electronic states determined by STS
are transport states because electron tunneling into unoccupied
molecular states results in the formation of negative
ion resonances. On the other hand, tunneling out of occupied
molecular states leads to the creation of positive ion
resonances. The same is true for 2PPE measurements, i.e.,
when unoccupied states are populated via an electron
transfer from the metal to the molecule (creating a negative ion
resonance) or when occupied molecular electronic states are ionized, creating a positive ion. Thus, in this case
we obtain the ionization potential (IP) or the electron affinity (EA) of the
respective molecular state. Hence, both STS and 2PPE
enable the determination of transport levels and consequently
transport gaps. In addition, in 2PPE an intramolecular
excitation, e.g., a direct HOMO to LUMO transition is also possible (the molecule
remains neutral), which is needed for exciton generation.
The energy required for this process is $E_{opt}$, the optical gap.
$E_{opt}$ is lower than the difference between the IP
and EA of the respective HOMO and
LUMO levels (transport gap, $E_{transp}$): $E_{transp} = IP - EA = E_{B} + E_{opt}$ with $E_{B}$ the exciton
binding energy \cite{Muntwiler2010, Knupfer2003}.
Note that the exciton binding energy of a molecule on a surface consists of correlations within the molecule and contributions by the metallic substrate (screening effects).
For the exciton, the
ionization potential is measured in 2PPE.

Combining our results described above, for 1ML FeTPyP on Au(111), we obtain a transport gap of 2.2 eV (EA(LUMO) = 3.21 eV; IP(HOMO) = 5.41 eV).
The excitonic state exhibits a binding energy ($E_{B}$) of 420 meV.  Thus, the optical gap is 1.78 eV.
Apart from the HOMO, we determined further lower lying occupied FeTPyP-derived states (OMO$_{2}$--OMO$_{4}$). Furthermore in the energy region of the unoccupied molecular states, the energetic position of the LUMO+1 and a higher lying state (UMO$_{3}$) have been identified. A further unoccupied molecular final state is found, which is located 0.68$\pm$0.07 eV above the vacuum level.

\section{Conclusion}
Deposition of Fe-tetra-pyridil-porphyrin (FeTPyP) molecules onto a Au(111) surface at room temperature leads to well-ordered large self-assembled islands as demonstrated by low-temperature scanning tunneling microscopy. Tunneling spectroscopy revealed the energies of negative ion resonances, commonly referred to as the LUMO and LUMO+1. At negative bias voltages, the resolution of positive ion resonances, or commonly referred to occupied molecular orbitals is more difficult due to the higher effective tunneling barrier. In contrast, UPS is ideally suited to determine the alignment of occupied states, thus being well complementary to STS measurements. In addition, our 2PPE data does not only confirm the energies of unoccupied molecular states, but is also capable of revealing intramolecular photoexcitations, which effectively reveal an optical gap. Comparison to the ionized electronic levels, we then determined an exciton binding energy of 420 meV.
The quantitative determination of the energetic position of
molecular electronic levels  with respect to the Fermi level of
an electrode is essential for the improvement and optimization of molecule-based
device performance.

\section{Acknowledgments}
A.S. and P.T. thank Elena Michel and Sonja Wieland for their
assistance and experimental support. They acknowledge funding by the German Research Foundation (DFG)
through collaborative research center SFB1249 "N-Heteropolycylces as Functional Materials" (project B06).
D.R., C.L., and K.J.F. acknowledge funding by the ERC Consolidator Grant "NanoSpin" and the
International Max Planck Research School on "Functional Interfaces in Physics and
Chemistry".

\section{References}

%\bibliographystyle{unsrt}
%\bibliographystyle{iopart-num}

%\bibliography{porphyrin_lit}

\end{document}